\documentclass[smallextended]{svjour3}
\usepackage{makeidx,graphics,subfigure}
\usepackage{graphicx}

\begin{document}
\title{Modulation Algorithms for Manipulating Nuclear Spin States}

\author{Boyang Liu \and
        Ming~Zhang  \and
        Hong-Yi Dai
}


\institute{Boyang Liu, Ming~Zhang\at
              Department of Automatic control, College of Mechatronic Engineering
and Automation, National University of Defense Technology,
Changsha, Hunan 410073, People's Republic of China\\
              Tel.: 0086-731-84573333\\
              Fax: 0086-731-84573323\\
              \email{zhangming@nudt.edu.cn}           %  \\
                      \and
        Hong-Yi Dai \at
        Department of Physics, College of Science, National University of Defense Technology,
Changsha, Hunan 410073, People's Republic of China\\
 }

\date{Received: date / Accepted: date}

\maketitle%

\begin{abstract}
%\boldmath
We exploit the impact of exact frequency modulation on transition time of steering nuclear spin states from a theoretical point of view. 1-stage and 2-stage Frequency-Amplitude-Phase modulation (FAPM) algorithms are proposed in contrast with 1-stage and 3-stage Amplitude-Phase modulation (APM) algorithms. The sufficient conditions are further presented for transiting nuclear spin states within the specified time by these four modulation algorithms. It is demonstrated that transition time performance can be significantly improved if exact frequency modulation is available. It is exemplified that the transition time scale with frequency modulation is about 1/4 of that without frequency modulation. It is also revealed in this research that the hybrid scheme of 1-stage FAPM and APM algorithms is better than all the four modulation algorithms. By simplifying the hybrid control algorithm, an effective hybrid modulation algorithm is also proposed to reduce computational burden.
\end{abstract}

%\begin{IEEEkeywords}
%Decision theory, Quantum mechanics, Bayesian methods, a priori
%knowledge
%\end{IEEEkeywords}

% \fi

\section{Introduction}

Dating from the birth of quantum theory, control of quantum systems is an important issue.
Quantum control theory has been developed ever since last century.
Recently, quantum information and quantum computation is the focus of the research\cite{Nielsen}.

In this paper, we will concentrate on nuclear magnetic resonance (NMR) systems\cite{Ernst},
which is one of the promising implementations of quantum information and quantum computation.
The existence of nuclear spin and its associated magnetism was first suggested by W. Pauli in 1924,
and the interaction of this nuclear magnetism with an external magnetic field was predicted
to result in a finite number of discrete energy levels known as the Zeeman structure.
In 1933, the first direct excitation of transitions between nuclear Zeeman levels was achieved by I. Rabi\cite{Rabi},
and the first NMR experiments were performed by F. Bloch and co-workers at Stanford University\cite{Bloch} and E. Purcell and co-workers at MIT\cite{Pucell} in 1945.
In 1950, E. Hahn discovered the spin echo\cite{Hahn}, thus opening the possibility of manipulating spin coherence.
It is not until 1995 that Di-Vincenzo first suggested the utilization of nuclear spins in quantum computation.
A great deal of contribution has been made for NMR approach to quantum information processing\cite{Nielsen}.
It has been demonstrated theoretically and experimentally that Amplitude and Phase modulation methods can be used for manipulating nuclear spin states\cite{Vandersypen}. Recently, quantum algorithms such as nonadiabatic holonomic quantum computation\cite{Feng}, quantum factorization\cite{Xu,Peng}, quantum simulation\cite{Feng2,Feng3,Du,Lu},
and optimal quantum control for $^{2}H$\cite{Wei} have been successfully realized by using NMR systems in experiments.
On the other hand, with the utilization of NMR devices, physical dynamics like quantum Zeno Effect\cite{Zheng}
and room temperature activation of methane\cite{Xu2}, new control systems like two-qubit coupling systems\cite{Yuan,Assemat}
and Multiple-spin coherence transfer systems\cite{Nimbalkar,Du2} and promising control schemes like Geometric Optimal Control\cite{Tosner,Bonnard}
also become easier to be studied or applied and have roused the attention of many researchers.

With the development of technology, exact frequency modulation methods will be utilized for steering nuclear spin states in the future.
In this paper, we explored what will happen
when frequencies of dynamical radio-frequency (RF) field are permitted to be adjusted exactly within an interval including resonance frequency.
In order to get some analytical expressions of modulation algorithms,
we focus on a simple but general NMR model,
in which a nuclear spin is modulated by a parameter adjustable electromagnetic field.
We discuss the influence of frequency adjustability in that NMR system by exploiting 4 kinds of modulation control algorithms:
3-stage Amplitude-Phase modulation (APM) algorithm,
1-stage APM algorithm, 2-stage Frequency-Amplitude-Phase modulation (FAPM) algorithm and 1-stage FAPM algorithm.
It is revealed that two kinds of FAPM control methods can both be utilized to remarkably improve transition time of nuclear spin states steering
as long as exact frequency modulation is available.
It is exemplified that transition time scale with frequency modulation is about $1/4$ of that without frequency modulation.
Further analysis also indicates that 1-stage FAPM method is not always better than 1-stage APM method in terms of transition time performance,
and this suggests that the hybrid modulation algorithm based on FAPM and APM methods is better than all the aforementioned algorithms
in terms of time performance and is an excellent candidate for three-parameter-modulation control methods.
A simplified hybrid algorithm is further proposed to reduce the computational burden without sacrificing time performance.
Furthermore, algorithms are also discussed from the viewpoint of optimal control\cite{Boscain,DAlessandro,Khaneja,Yuan2}.
In terms of time performance, none of the aforementioned control algorithms is the optimal scheme for state transition.
However, it is beneficial to consider a sub-optimal problem.
Under the constraints of forcing target state achieved with no error and limiting control methods within the aforementioned algorithms,
the sub-optimal control scheme and its transition time are given analytically.

The rest of this paper is organized as follows.
Problem descriptions are presented in Sect. 2.
Four kinds of modulation algorithms are presented in Sect. 3
with their sufficient conditions for steering nuclear spin state within a specified time investigated.
Sect. 4 gives further analysis and presents two hybrid algorithms. This paper concludes with Sect. 5.

\section{Problem Description}

A controlled nuclear spin system is described by
\begin{equation}
\label{2-1}
\frac{d}{dt}|\psi(t)\rangle=\frac{i\gamma}{\hbar}[B_{0}H_0+B_{1}H_1(t)]|\psi(t)\rangle
\end{equation}
Here $B_0$ and $B_1$ represent static and dynamical radio-frequency (RF) field,
respectively, $\omega_{rf}$¦Ø and $\varphi$ are the frequency and phase of dynamical field,
and $\gamma$ is the gyromagnetic ratio of nucleus.
$H_0$ denotes the constant part of the Hamiltonian decided by NMR device itself,
while $H_1(t)$ represents Hamiltonian¡¯s variable part,
which we can deploy arbitrarily.
In this paper, we set
\begin{equation}
\label{2-1-1}
H_0=S_{z}=\frac{1}{2}\sigma_{z}=\frac{1}{2}\left(\begin{array}{cc}
                                                   1 & 0 \\
                                                   0 & -1
                                                 \end{array}
\right)
\end{equation}
  and
\begin{equation}
\label{2-1-2}
  H_1(t)=S_{x}\cos{(\omega_{rf}{t}+\phi)}-S_{y}\sin({\omega_{rf}{t}}+\phi)
\end{equation}
 with
$S_{x}=\frac{1}{2}\sigma_{x}=\frac{1}{2}\left(\begin{array}{cc}
                                                   1 & 0 \\
                                                   0 & -1
                                                 \end{array}\right)$,
and
$S_{y}=\frac{1}{2}\sigma_{y}=\frac{1}{2}\left(\begin{array}{cc}
                                                   0 & -i \\
                                                   i & 0
                                                 \end{array}\right)$.
Eq. (2) reflects a static magnetic field along some Z axis and Eq. (3)
is used to describe the adjustable magnetic field
in the normal space of Z axis\cite{Vandersypen}.
Furthermore, if an error between final state
and target state is acceptable,
the Hamiltonian given in Eq. (3) will lead to
an energy-error optimal control
within some fixed time\cite{Yuan2}.
As a result, $H_1$ (t) is given in such a form.

Denoting $\omega_{0}=\frac{\gamma B_{0}}{\hbar}$
and
 $\omega_{1}=\frac{\gamma B_{1}}{\hbar}$,
 with Eqs. (2) and (3),
 Eq. (1) can be further rewritten as
 \begin{equation}
 \label{2-1-3}
 \frac{d}{dt}|\psi(t)\rangle=i\{\omega_0 S_z +\omega _1[S_{x}cos(\omega_{rf}+\varphi)-S_ysin(\omega_{rf}+\varphi)]\}|\psi(t)\rangle
 \end{equation}
where $\omega_1$, $\omega_{rf}$ and $\varphi$ are adjustable control parameters. Our control goal is to steer the nuclear spin system from an arbitrary initial state $|\psi_0\rangle$ to another arbitrary target state $|\psi_f\rangle$ by adjusting $\omega_1$, $\omega_{rf}$ and $\varphi$. Considering the question whether or not one can manipulate the nuclear spin system from $\psi_0$ to $\psi_{rf}$ within the specified time $T$, we will also give some discussions about the influence of exact frequency modulation on time performance.

Suppose that the permissible set is given as
$U=\{\omega_{1}[S_{x}\cos({\omega_{rf}{t}+\varphi})-S_{y}\sin({\omega_{rf}{t}+\varphi})]:
\omega_{1}\in{S_{\omega_{1}}}\subseteq{R^{+}},\omega_{rf}\in{S_{\omega_{rf}}}\subseteq{R^{+}}, \varphi\in{R}\}$.
In this paper, 4 kinds of parameter modulation control algorithms will be explored:

(1)   $3$-stage APM algorithm:

Control Hamiltonian is given by
\begin{equation}
\label{Hc1}
H^{1}_c(t)=\left\{\begin{array}{ll}
0&\  if \  t\in[t_0,t_1)\\
\omega_1[S_x\cos[\omega_0(t-t_1)]-S_y\sin[\omega_0(t-t_1)]]&\  if \  t\in[t_1,t_2)\\
0&\  if \  t\in[t_2,t_f)
\end{array}\right.
\end{equation}
where $\omega_1$, $t_1$, $t_2$ and $t_f$ are designed parameters to be chosen with $I_{\omega_1}=[0,\omega^{max}_{1}]$ and phase
$\varphi=-\omega_0\cdot{t_1}$.

(2)     $1$-stage APM algorithm:

When $t\in[t_0,t_f)$, control Hamiltonian is given by
\begin{equation}
\label{Hc2}
H^{2}_c(t)=
\omega_1[S_x\cos[\omega_0(t-t_0)+\varphi_1]-S_y\sin[\omega_0(t-t_0)+\varphi_1]]
\end{equation}
where $\omega_1$, $\varphi_1$ and $t_f$ are designed parameters to be chosen with $I_{\omega_1}=[0,\omega^{max}_{1}]$ and phase
$\varphi=\varphi_1-\omega_0\cdot{t_0}$.

(3)  $2$-stage FAPM algorithm:

Control Hamiltonian is given by
\begin{equation}
\label{Hc3}
H^{3}_c(t)=\left\{\begin{array}{ll}
0&\  if \  t\in[t_0,t_1)\\
\omega_1[S_x\cos[\omega_{rf}(t-t_1)]-S_y\sin[\omega_{rf}(t-t_1)]]&\  if \  t\in[t_1,t_f)
\end{array}\right.
\end{equation}
where $\omega_1$, $\omega_{rf}$, $t_1$, and $t_f$ are designed parameters to be chosen with $I_{\omega_1}=[0,\omega^{max}_{1}]$ and phase
$\varphi=-\omega_{rf}\cdot{t_1}$ and $I_{\omega_{rf}}=[\omega_{0}-\omega^{-}_{b},\omega_0+\omega^{+}_{b}]$.

(4)   $1$-stage FAPM algorithm:

When $t\in[t_0,t_f)$, control Hamiltonian is given by
\begin{equation}
\label{Hc4}
H^{4}_c(t)
=\omega_1\{S_x\cos[\omega_{rf}(t-t_0)+\varphi_1]-S_y\sin[\omega_{rf}(t-t_0)+\varphi_1]\}
\end{equation}
where $\omega_1$, $\omega_{rf}$, $\varphi_1$, and $t_f$ are designed parameters to be chosen with $I_{\omega_1}=[0,\omega^{max}_{1}]$ and phase
$\varphi=\varphi_1-\omega_{rf}\cdot{t_0}$ and $I_{\omega_{rf}}=[\omega_{0}-\omega^{-}_{b},\omega_0+\omega^{+}_{b}]$.

By studying all these four modulation algorithms, we will further exploit the impact of exact frequency modulation on transition time for  manipulating
nuclear spin states.

\section{Modulation Algorithms}

For the purpose of simplicity, we introduce some notations in advance:
$|\psi_0\rangle=\cos{\frac{\theta_{0}}{2}}|\uparrow\rangle+e^{i\phi_{0}}\sin{\frac{\theta_{0}}{2}}|\downarrow\rangle$ and
$|\psi_f\rangle=\cos{\frac{\theta_{f}}{2}}|\uparrow\rangle+e^{i\phi_{f}}\sin{\frac{\theta_{f}}{2}}|\downarrow\rangle$
where $0\leq{\theta_{0}},{\theta_{f}}\leq\pi$ and $0\leq{\phi_{0}},{\phi_{f}}<2\pi$.

\subsection{$3$-stage APM Algorithm}
In this subsection, we will give a $3$-stage APM algorithm to design control Hamiltonian
(\ref{Hc1})
 to steer nuclear spin state from $|\psi_0\rangle$ to $|\psi_f\rangle$ as follows:

Choose
\begin{equation}
\label{aomga1}
\omega^{(1)}_1=\omega^{\max}_{1},
\end{equation}
\begin{equation}
\label{at1}
t^{(1)}_1=\frac{\phi_0+\frac{3\pi}{2}}{\omega_0}+t_0,
\end{equation}
\begin{equation}
\label{at2}
t^{(1)}_2=\frac{4\pi+\theta_f-\theta_0}{\omega^{\max}_1}+t^{(1)}_1
\end{equation}
and
\begin{equation}
\label{atf}
t^{(1)}_f=\frac{2k_1\pi+\frac{\pi}{2}-\phi_f}{\omega_0}-\frac{4\pi+\theta_f-\theta_0}{\omega^{\max}_1}+t^{(1)}_2
\end{equation}
where $k_1$ is such an integer that
$k_1\geq\frac{(4\pi+\theta_f-\theta_0)\omega_0}{2\pi\omega^{\max}_{1}}-\frac{1}{4}+\frac{\phi_f}{2\pi}$.

Therefore,   $t^{(1)}_f-t_0=\frac{2k_1\pi+{2\pi}-\phi_f+\phi_0}{\omega_0}$.
According to Lemma 2 given in the appendix, with $\tau=t_1^{(1)}$ and $t=t_2^{(1)}$, we have that
\begin{equation}
\label{psitf}
|\psi(t^{(1)}_f)\rangle=e^{i\omega_0(t^{(1)}_f-t^{(1)}_1){S_z}}exp\{i\omega_1(t^{(1)}_2-t^{(1)}_1)S_x\}e^{i\omega_0(t^{(1)}_1-t_0){S_z}}|\psi(t_0)\rangle.
\end{equation}
This implies
$|\psi(t^{(1)}_f)\rangle=\cos{\frac{\theta_{f}}{2}}|\uparrow\rangle+e^{i\phi_{f}}\sin{\frac{\theta_{f}}{2}}|\downarrow\rangle$
according to Lemma 4.

Notice that one can choose $t^{(1)}_1=\frac{\phi_0-\frac{\pi}{2}}{\omega_0}+t_0$ if $\phi_0-\frac{\pi}{2}\geq0$, and select
$t^{(1)}_2=\frac{\theta_f-\theta_0}{\omega^{\max}_1}+t^{(1)}_1$ if $\theta_f-\theta_0\geq0$.
Therefore the aforementioned algorithm can be further improved in terms of time performance according to the initial and target states as follows:

Choose
\begin{equation}
\label{at1a}
t^{(1)}_1=\left\{\begin{array}{ll}
\frac{\phi_0+\frac{3\pi}{2}}{\omega_0}+t_0&\  if \  \phi_0-\frac{\pi}{2}<0\\
\frac{\phi_0-\frac{\pi}{2}}{\omega_0}+t_0&\  if \  \phi_0-\frac{\pi}{2}\geq0
\end{array}\right.,
\end{equation}
%and
\begin{equation}
\label{at2a}
t^{(1)}_2=\left\{\begin{array}{ll}
\frac{4\pi+\theta_f-\theta_0}{\omega^{\max}_1}+t^{(1)}_1&\  if \  \theta_f-\theta_0<0\\
\frac{\theta_f-\theta_0}{\omega^{\max}_1}+t^{(1)}_1&\  if \  \theta_f-\theta_0\geq0
\end{array}\right.
\end{equation}
and
\begin{equation}
\label{atfa}
t^{(1)}_f=\left\{\begin{array}{ll}
\frac{2k_1\pi+\frac{\pi}{2}-\phi_f}{\omega_0}-\frac{4\pi+\theta_f-\theta_0}{\omega^{\max}_1}+t^{(1)}_2&\  if \  \theta_f-\theta_0<0\\
\frac{2k_1\pi+\frac{\pi}{2}-\phi_f}{\omega_0}-\frac{\theta_f-\theta_0}{\omega^{\max}_1}+t^{(1)}_2&\  if \  \theta_f-\theta_0\geq0
\end{array}\right.
\end{equation}
where $k_1$ is given by
\begin{equation}
\label{ka1}
k_{1}=\left\{\begin{array}{ll}
\min\{k\in{Z^{+}}|k\geq\frac{(4\pi+\theta_f-\theta_0)\omega_0}{2\pi{\omega^{max}_1}}-\frac{1}{4}+\frac{\phi_{f}}{{2\pi}}\}&\  if \  \theta_f-\theta_0<0 \\
\min\{k\in{Z^{+}}|k\geq\frac{(\theta_f-\theta_0)\omega_0}{2\pi{\omega^{max}_1}}-\frac{1}{4}+\frac{\phi_{f}}{{2\pi}}\}&\  if \  \theta_f-\theta_0\geq0
\end{array}\right..
\end{equation}

Therefore the  transition time $t^{(1)}_{f}-t_0$ is given by
\begin{equation}
\label{1tft0}
t^{(1)}_f-t_0=\left\{\begin{array}{ll}
\frac{2k_1\pi+2\pi-\phi_f+\phi_0}{\omega_0}&\  if \  \phi_0-\frac{\pi}{2}<0\\
\frac{2k_1\pi-\phi_f+\phi_0}{\omega_0}&\  if \  \phi_0-\frac{\pi}{2}\geq0
\end{array}\right..
\end{equation}

\textbf{Theorem 1}: For the controlled nuclear system given by Eq. (\ref{2-1-3}) with $|\omega_1|\leq\omega^{\max}_1$, there exists a $3$-stage APM algorithm
to steer nuclear spin system from an arbitrary initial state
to another arbitrary target state
within the   specified time $T>0$  if
\begin{equation}
\label{theorem1}\frac{4\pi}{\omega^{\max}_1}+\frac{7.5\pi}{\omega_0}\leq{T}.
\end{equation}

\textbf{Proof:}   To uniformly  estimate upper bound of the transition time $t^{(1)}_{f}-t_0$ for  $3$-stage APM algorithm, we have from Eq. (\ref{ka1})
that
\begin{equation}
\label{aka1}
k_{1}<k^{*}_{1}=\left\{\begin{array}{ll}
\min\{k\in{Z^{+}}|k\geq\frac{2\omega_0}{{\omega^{max}_1}}+\frac{3}{4}\}&\  if \  \theta_f-\theta_0<0 \\
\min\{k\in{Z^{+}}|k\geq\frac{\omega_0}{2{\omega^{max}_1}}+\frac{3}{4}\}&\  if \  \theta_f-\theta_0\geq0
\end{array}\right..
\end{equation}
Thus, transition time $t^{(1)}_f-t_0$ satisfies the following inequality
\begin{equation}
\label{2-2-12}
t^{(1)}_f-t_0\leq\frac{4\pi}{\omega^{\max}_1}+\frac{7.5\pi}{\omega_0}
\end{equation}
 for any pair of initial and target states.

Therefore, the sufficient condition for there exists a $3$-stage APM algorithm to steer nuclear spin system from an arbitrary initial state
to another arbitrary target state within the   specified time $T>0$ is  given by Eq. (\ref{theorem1}).

\textbf{Remark 1}: For example, $\omega^{max}_1=50kHz$ and $\omega_{0}=500MHz$ hold typically for nucleons $^{1}H$. This implies that
\begin{equation}
\label{remark1}
\frac{4\pi}{\omega^{\max}_1}+\frac{7.5\pi}{\omega_0}=\frac{4.00075\pi}{\omega^{max}_1}\approx\frac{4\pi}{\omega^{max}_1}=2.512\times10^{-4}s.
\end{equation}
 Therefore, one can steer nucleons $^{1}H$ from an arbitrary initial state to another arbitrary state within about $2.512\times10^{-4}s$ by using
 $1$-stage APM algorithm.

\subsection{$1$-stage APM algorithm}
In this subsection, we will construct a $1$-stage APM algorithm to design control Hamiltonian
(\ref{Hc2})
 to steer nuclear spin state from $|\psi_0\rangle$ to $|\psi_f\rangle$ as follows:

Choose
\begin{equation}
\label{apa1}
\varphi^{(2)}_1=\frac{\pi}{2}-\phi_0,
\end{equation}
\begin{equation}
\label{apa2}
\omega^{(2)}_1=\frac{(4\pi+\theta_f-\theta_0)\omega_0}{\phi^{(2)}_{k}}
\end{equation}
and
\begin{equation}
\label{aptf1}
t^{(2)}_f=\frac{\phi^{(2)}_{k}}{\omega_0}+t_0
\end{equation}
where
\begin{equation}
\label{phik2}
\phi^{(2)}_{k}={2k_{2}\pi+\phi_0-\phi_f}
\end{equation}
and $k_{2}$ is such an integer that $k_{2}\geq\frac{(4\pi+\theta_f-\theta_0)\omega_0}{2\pi\omega^{\max}_{1}}+\frac{\phi_f-\phi_0}{2\pi}$.

From $\omega_{0}{\cdot}(t^{(2)}_{f}-t_0)+\varphi^{(2)}=2k_{2}\pi-\phi_f+\frac{\pi}{2}$ and Lemma 2 with $t=t^{(2)}_{f}$, $\tau=t_0$ and
$\varphi_1=\varphi^{(2)}_1$, we have
\begin{equation}
\label{2-2-7}
|\psi(t^{(2)}_{f})\rangle=e^{i[\omega_{0}(t^{(2)}_{f}-t_0)+\varphi^{(2)}_1]S_{z}}exp\{{i\omega_1{(t^{(2)}_{f}-t_0)}}S_{x}\}e^{-i{\varphi^{(2)}_1}S_{z}}|\psi(t_0)\rangle.
\end{equation}
Therefore, we conclude from Lemma 4 and
$|\psi(t_0)\rangle=\cos{\frac{\theta_{0}}{2}}|\uparrow\rangle+e^{i\phi_{0}}\sin{\frac{\theta_{0}}{2}}|\downarrow\rangle$
that
\begin{equation}
\label{2-2-9}
|\psi(t^{(2)}_{f})\rangle
=\cos{\frac{\theta_{f}}{2}}|\uparrow\rangle+e^{i\phi_{f}}\sin{\frac{\theta_{f}}{2}}|\downarrow\rangle.
\end{equation}

Noticing that ${\omega^{(2)}_1}$ in Eq. (\ref{apa2}) can be modified when $\theta_f-\theta_0\geq0$, $1$-stage APM  algorithm can be improved in terms of time performance according to the initial and target states as follows:

Change  $\omega^{(2)}_{1}$ and $k_2$ into
\begin{equation}
\label{oapomega1}\omega^{(2)}_{1}=
\left\{\begin{array}{ll}
\frac{(\theta_f-\theta_0)\omega_0}{2k_{2}\pi+\phi_0-\phi_f}&if \  \theta_f-\theta_0\geq0\\
\frac{(4\pi+\theta_f-\theta_0)\omega_0}{2k_{2}\pi+\phi_0-\phi_f}&if \  \theta_f-\theta_0<0
\end{array}\right.
\end{equation}
 and
\begin{equation}
\label{kap1}
k_{2}=\left\{\begin{array}{ll}
\min\{k\in{Z^{+}}|k\geq\frac{(4\pi+\theta_f-\theta_0)\omega_0}{2\pi{\omega^{max}_1}}+\frac{\phi_{f}-\phi_0}{{2\pi}}\}&\  if \  \theta_f-\theta_0<0 \\
\min\{k\in{Z^{+}}|k\geq\frac{(\theta_f-\theta_0)\omega_0}{2\pi{\omega^{max}_1}}+\frac{\phi_{f}-\phi_0}{{2\pi}}\}&\  if \  \theta_f-\theta_0\geq0
\end{array}\right.,
\end{equation}
respectively.
To uniformly  estimate upper bound of the transition time $t^{(2)}_{f}-t_0$ for $1$-stage APM algorithm, we have
\begin{equation}
\label{2-2-10}
k_{2}{\leq}k^{*}_{2}=\left\{\begin{array}{ll}
\min\{k\in{Z^{+}}|k\geq\frac{2\omega_0}{{\omega^{\max}_1}}+1\}&\  if \  \theta_f-\theta_0<0\\
\min\{k\in{Z^{+}}|k\geq\frac{\omega_0}{2{\omega^{\max}_1}}+1\}&\  if \  \theta_f-\theta_0\geq0
\end{array}\right..
\end{equation}
Thus,
\begin{equation}
\label{2-2-11a} \phi^{(2)}_{k}=2k_{2}\pi+\phi_0-\phi_f\leq2(\frac{2\omega_0}{\omega^{\max}_{1}}+2)\pi+2\pi,
\end{equation}
and transition time $t^{(2)}_f-t_0$ satisfies the following inequality
\begin{equation}
\label{2-2-12a}
t^{(2)}_f-t_0=\frac{\phi^{(2)}_{k}}{\omega_0}\leq\frac{4\pi}{\omega^{\max}_1}+\frac{6\pi}{\omega_0}
\end{equation}
 for any pair of initial and target states.

Therefore, we have the following theorem:

\textbf{Theorem 2}: For the controlled nuclear spin system given by Eq. (\ref{2-1-3}) with $|\omega_1|\leq\omega^{\max}_1$, there exists a $1$-stage APM
algorithm   to steer the nuclear spin system from an arbitrary initial state
to another arbitrary target state
within the   specified time $T>0$  if
\begin{equation}
\label{theorem2}
\frac{4\pi}{\omega^{\max}_1}+\frac{6\pi}{\omega_0}\leq{T}.
\end{equation}

\textbf{Remark 2}: For example, $\omega^{max}_1=50kHz$ and $\omega_{0}=500MHz$ hold typically for nucleons $^{1}H$. This implies that
\begin{equation}
\label{remark2}
\frac{4\pi}{\omega^{\max}_1}+\frac{6\pi}{\omega_0}=\frac{4.0006\pi}{\omega^{max}_1}\approx\frac{4\pi}{\omega^{max}_1}=2.512\times10^{-4}s.
 \end{equation}
 Therefore, one can steer nucleons $^{1}H$ from an arbitrary initial state to another arbitrary state within about $2.512\times10^{-4}s$ by using
 $1$-stage APM algorithm.

\subsection{$2$-stage FAPM algorithm}

In this subsection, we will give a $2$-stage FAPM algorithm to design control Hamiltonian
(\ref{Hc3})
 to steer nuclear spin state from $|\psi_0\rangle$ to $|\psi_f\rangle$ as follows:

Choose
\begin{equation}
\label{fa1}
\omega^{(3)}_{1}={\frac{\omega_0\pi\sin\frac{\theta_0+\theta_f}{2}}{\phi^{(3)}_{k}}},
\end{equation}
\begin{equation}
\label{fa2}
\omega^{(3)}_{rf}=\frac{(2k_{3}\pi-\phi_{f})\omega_0}{\phi^{(3)}_{k}},
\end{equation}
\begin{equation}
\label{fat1}t^{(3)}_1=\frac{\phi_{0}}{\omega_0}+t_0
\end{equation}
and
\begin{equation}
\label{fatf}t^{(3)}_f=\frac{\phi^{(3)}_{k}}{\omega_0}+t^{(3)}_1
\end{equation}
where
\begin{equation}
\label{faphi}
\phi^{(3)}_{k}=2k_{3}\pi-\phi_{f}+\pi{\cos\frac{\theta_{0}+\theta_{f}}{2}}
\end{equation}
and $k_{3}$ is such an  integer that
\begin{equation}
\label{fakfa}
k_{3}=\min{\{k\in{Z^{+}}|k\geq{R_{\omega}}+R_{B1}\}}
\end{equation}
with
\begin{equation}
\label{Rowema}
R_{\omega}=\max\{\frac{\omega_0\sin\frac{\theta_0+\theta_f}{2}}{2\omega^{max}_1},\frac{\omega_0|\cos\frac{\theta_0+\theta_f}{2}|}{2\min\{\omega^{+}_{b},\omega^{-}_{b}\}}\}
\end{equation}
and
\begin{equation}
\label{Rb1}
R_{B1}=\frac{\phi_f-\pi\cos\frac{\theta_0+\theta_f}{2}}{2\pi}.
\end{equation}
%
%For arbitrary $\theta_0$, $\phi_0$, $\theta_f$ and  $\phi_f$, one can always find a sufficiently large integer $k_{3}\in{Z^{+}}$ so that
%$\phi^{3}_{k}>0$,  $\omega_1\in[0,\omega^{\max}_{1}]$ and  $\omega_{rf}\in[\omega_0-\omega^{-}_b,\omega_0+\omega^{+}_b]$ hold.
Therefore,  the transition time $t^{(3)}_f-t_0$ is given by
\begin{equation}
\label{tft03}t^{(3)}_f-t_0=\frac{2k_{3}\pi-\phi_{f}+\phi_{0}+\pi{\cos\frac{\theta_{0}+\theta_{f}}{2}}}{\omega_0}.
\end{equation}

Applying Lemma 3 with $\omega_{rf}=\omega^{(3)}_{rf}$, $t=t^{(3)}_{f}$, $\tau=t^{(3)}_1$ and $\varphi_1=0$, we have
\begin{equation}
\label{Lemma3-2a}
|\psi(t^{(3)}_f)\rangle=
e^{i[\omega^{(3)}_{rf}(t^{(3)}_f-t^{(3)}_1)]S_z}exp\{iR_u(t^{(3)}_{f}-t^{(3)}_1)S_{\theta_u}\}e^{i\omega_0(t^{(3)}_1-t_0){S_z}}|\psi(t_0)\rangle.
\end{equation}

 Noticing that $R_{u}=\frac{\omega_0\pi}{\phi^{(3)}_{k}}$, ${R_{u}{(t^{(3)}_{f}-t_1)}}=\pi$,
 $\omega^{(3)}_{rf}{(t^{(3)}_{f}-t^{(3)}_1)}=2k_{3}\pi-\phi_{f}$,
and $\theta_{u}=\frac{\theta_{0}+\theta_{f}}{2}$, we conclude from Lemma 5  that
\begin{equation}
\label{2FAPMtf}|\psi(t^{(3)}_{f})\rangle
=\cos{\frac{\theta_{f}}{2}}|\uparrow\rangle+e^{i\phi_{f}}\sin{\frac{\theta_{f}}{2}}|\downarrow\rangle.
\end{equation}

Furthermore, we have the following theorem:

\textbf{Theorem 3}: For the controlled nuclear spin system given by Eq. (\ref{2-1-3}) with $|\omega_1|\leq\omega^{\max}_1$ and
$\omega_{0}-\omega^{-}_{b}\leq\omega_{rf}\leq\omega_0+\omega^{+}_{b}$, there exists a $2$-stage FAPM algorithm   to steer nuclear spin system from an
arbitrary initial state
to another arbitrary target state
within the   specified time $T>0$  if
\begin{equation}
\label{theorem3}
\frac{\pi}{\min\{\omega^{max}_1,\omega^{+}_{b},\omega^{-}_{b}\}}+\frac{8\pi}{\omega_0}\leq{T}.
\end{equation}

\textbf{Proof:}  To uniformly  estimate upper bound of the transition time $t^{(3)}_{f}-t_0$ for $2$-stage FAPM algorithm, we have
\begin{equation}
\label{faRomega}
R_{\omega}\leq\frac{\omega_0}{2\min\{\omega^{max}_1,\omega^{+}_{b},\omega^{-}_{b}\}}=R^{*}_{\omega}
\end{equation}
and
\begin{equation}
\label{faRb}
R_{B1}\leq\frac{3}{2}.
\end{equation}
Thus
\begin{equation}
\label{fakb}
k_{3}{\leq}R^{*}_{\omega}+\frac{3}{2}+1
\end{equation}
since $k_3\in{Z^{+}}$. Therefore,
\begin{equation}
\label{faphik}
\phi^{(3)}_{k}\leq2[\frac{\omega_0}{2\min\{\omega^{max}_1,\omega^{+}_{b},\omega^{-}_{b}\}}+\frac{5}{2}]\pi+\pi,
\end{equation}
and the transition time $t^{(3)}_f-t_0$ satisfies the following inequality
\begin{equation}
\label{fatf3}
t^{(3)}_f-t_0\leq\frac{\pi}{\min\{\omega^{max}_1,\omega^{+}_{b},\omega^{-}_{b}\}}+\frac{8\pi}{\omega_0}
\end{equation}
for any pair of initial and target states.
Therefore,  the sufficient condition for there exists a $2$-stage FAPM algorithm  to steer nuclear spin system from an arbitrary initial state
to another arbitrary target state
within the   specified time $T>0$ is  given by Eq. (\ref{theorem3}).

\textbf{Remark 3}: If $\omega_0=50MHz$ and $\min\{\omega^{+}_{b},\omega^{-}_{b}\}\geq\omega^{max}_1=50kHz$, then
\begin{equation}
\label{remark3}
\frac{\pi}{\min\{\omega^{max}_1,\omega^{+}_{b},\omega^{-}_{b}\}}+\frac{8\pi}{\omega_0}\leq\frac{1.0008\pi}{\omega^{max}_1}\approx\frac{\pi}{\omega^{max}_1}=6.28\times10^{-5}s.
\end{equation}
 Therefore, one can steer nucleons $^{1}H$ from an arbitrary initial state to another arbitrary state within about $6.28\times10^{-5}s$ by using $2$-stage
 FAPM algorithm.

\subsection{$1$-stage FAPM algorithm}

In this subsection, we will describe a $1$-stage FAPM algorithm to design control Hamiltonian
(\ref{Hc4})
 to steer nuclear spin state from $|\psi_0\rangle$ to $|\psi_f\rangle$ as follows:

Choose
\begin{equation}
\label{fap1}
\varphi_{1}^{(4)}=-\phi_0,
\end{equation}
\begin{equation}
\label{fap2}
\omega^{(4)}_{1}=
{\frac{\omega_0\pi\sin\frac{\theta_0+\theta_f}{2}}{\phi^{(4)}_{k}}},
\end{equation}
\begin{equation}
\label{fap3}
\omega^{(4)}_{rf}=\frac{(2k_{4}\pi-\phi_{f}+\phi_0)\omega_0}{\phi^{(4)}_{k}}
\end{equation}
and
\begin{equation}
\label{faptf}t^{(4)}_f=\frac{\phi^{(4)}_{k}}{\omega_0}+t_0
\end{equation}
where
\begin{equation}
\label{fapphi}
\phi^{(4)}_{k}=2k_{4}\pi-\phi_{f}+\phi_{0}+\pi{\cos\frac{\theta_{0}+\theta_{f}}{2}}
\end{equation}
and $k_{4}$ is such an  integer that
\begin{equation}
\label{fapkfap}
k_{4}=\min\{k\in{Z^{+}}|k\geq{R_{\omega}}+R_{B2}\}
\end{equation}
with \begin{equation}
\label{Rb2}
R_{B2}=\frac{\phi_f-\phi_0-\pi\cos\frac{\theta_0+\theta_f}{2}}{2\pi},
\end{equation}
and $R_{\omega}$ is given by Eq. (\ref{Rowema})

%For arbitrary $\theta_0$, $\phi_0$, $\theta_f$ and  $\phi_f$, one can always find sufficiently large $k_{4}\in{Z^{+}}$ so that
%$\phi^{(4)}_{k}>0$,  $\omega^{(4)}_1\in[0,\omega^{\max}_{1}]$ and  $\omega^{(4)}_{rf}\in[\omega_0-\omega^{-}_b,\omega_0+\omega^{+}_b]$ hold.

From Lemma 3 with $\omega_{rf}=\omega^{(4)}_{rf}$, $\varphi_1=\varphi^{(4)}_1$, $\tau=t_0$ and $t=t^{(4)}_f$, we have
\begin{equation}
\label{4-7c}
|\psi(t^{(4)}_{f})\rangle=e^{i[\omega^{(4)}_{rf}{(t^{(4)}_{f}-t_0)}+\varphi^{(4)}_1]S_{z}}exp\{{iR_{u}{(t^{(4)}_{f}-t_0)}}S_{\theta_{u}}\}e^{-i{\varphi^{(4)}_1}S_{z}}|\psi(t_0)\rangle.
\end{equation}

 After some calculations,  we conclude
from Lemma 5  that
\begin{equation}
\label{1fapmtf}
|\psi(t^{(4)}_{f})\rangle
=\cos{\frac{\theta_{f}}{2}}|\uparrow\rangle+e^{i\phi_{f}}\sin{\frac{\theta_{f}}{2}}|\downarrow\rangle.
\end{equation}

Furthermore, we have the following Theorem:

\textbf{Theorem 4}: For the controlled nuclear spin system given by Eq. (\ref{2-1-3}) with $|\omega_1|\leq\omega^{\max}_1$ and
$\omega_{0}-\omega^{-}_{b}\leq\omega_{rf}\leq\omega_0+\omega^{+}_{b}$, there exists a $1$-stage FAPM  algorithm  to steer nuclear spin system from an
arbitrary initial state
to another arbitrary target state
within the   specified time $T>0$  if
\begin{equation}
\label{theorem4}
\frac{\pi}{\min\{\omega^{max}_1,\omega^{+}_{b},\omega^{-}_{b}\}}+\frac{8\pi}{\omega_0}\leq{T}.
\end{equation}

\textbf{Proof:}  To uniformly  estimate upper bound of the transition time $t^{(4)}_{f}-t_0$ for $1$-stage FAPM algorithm, we have
\begin{equation}
\label{fapR4-1-1}
R_{\omega}\leq\frac{\omega_0}{2\min\{\omega^{max}_1,\omega^{+}_{b},\omega^{-}_{b}\}}=R^{*}_{\omega}
\end{equation}
and
\begin{equation}
\label{fapR4-1-2}
R_{B}\leq\frac{3}{2}.
\end{equation}
Thus,
\begin{equation}
\label{fapR4-1-3}
k_{4}{\leq}R^{*}_{\omega}+\frac{3}{2}+1
\end{equation}
since $k_4\in{Z^{+}}$.

Thus,
\begin{equation}
\label{fapR4-1-4}
\phi^{(4)}_{k}\leq2[\frac{\omega_0}{2\min\{\omega^{max}_1,\omega^{+}_{b},\omega^{-}_{b}\}}+\frac{5}{2}]\pi+3\pi,
\end{equation}
and transition time $t^{(4)}_f-t_0$ satisfies the following inequality
\begin{equation}
\label{C4-1}
t^{(4)}_f-t_0\leq\frac{\pi}{\min\{\omega^{max}_1,\omega^{+}_{b},\omega^{-}_{b}\}}+\frac{8\pi}{\omega_0}
\end{equation}
for any pair of initial and target states.

Therefore,  the sufficient condition for there exists a $1$-stage FAPM algorithm  to steer nuclear spin system from an arbitrary initial state
to another arbitrary target state
within the   specified time $T>0$ is  given by Eq. (\ref{theorem4}).

%b. When exact frequency modulation methods are available, we have
%\begin{equation}
%\label{RC4-1}
%{T_{fap}}<(1+\frac{8}{k_0})\frac{\pi}{\min\{\omega^{max}_1,\omega^{+}_{b},\omega^{-}_{b}\}}
%\end{equation}
%where $k_0\in{Z^{+}}$ and
%$(k_0+1)>\frac{{\omega_{0}}}{\min\{\omega^{max}_1,\omega^{+}_{b},\omega^{-}_{b}\}}{\geq}k_0$.
%
\textbf{Remark 4}: If $\omega_0=50MHz$ and $\min\{\omega^{+}_{b},\omega^{-}_{b}\}\geq\omega^{max}_1=50kHz$, then
\begin{equation}
\label{remark4}
\frac{\pi}{\min\{\omega^{max}_1,\omega^{+}_{b},\omega^{-}_{b}\}}+\frac{8\pi}{\omega_0}\leq\frac{1.0008\pi}{\omega^{max}_1}\approx\frac{\pi}{\omega^{max}_1}=6.28\times10^{-5}s.
\end{equation}
  Therefore, one can steer nucleons $^{1}H$ from an arbitrary initial state to another arbitrary state within about $6.28\times10^{-5}s$ by using
  $1$-stage FAPM algorithm.

\section{Discussions}
Based on 4 modulation algorithms in Sect. 3, we will do some further discussions in this section. Through the whole section, we assume for the
feasibility that $\min\{\omega^{+}_{b},\omega^{-}_{b}\}\geq\omega^{max}_1=50kHz$ and $\omega_{0}=500MHz$.

1. By comparing Eq. (\ref{ka1}) with Eq. (\ref{kap1}), we have $k_1\geq{k_2}$ if $\phi_0-\frac{\pi}{2}\geq{0}$ and $k_1+1\geq{k_2}$ if
$\phi_0-\frac{\pi}{2}<{0}$. This implies that
$t^{(2)}_{f}-t_0\leq{t^{(1)}_{f}}-t_0$  for any pair of initial and target states. In other words, $1$-stage APM algorithm is  better than  $3$-stage APM
algorithm
in terms of time performance. It is also emphasized that $|t^{(1)}_{f}-t^{(2)}_{f}|\leq\frac{2\pi}{\omega_0}$ holds for  any pair of initial and target
states.

2. Since $R_{B2}\leq{R_{B1}}$, $k_3\geq{k_4}$ holds for  for any pair of initial and target states, we conclude from Sect. 3.3 and 3.4 that
$t^{(4)}_{f}-t_0\leq{t^{(3)}_{f}}-t_0$ holds for  for any pair of initial and target states. That is, $1$-stage FAPM algorithm is  better than  $2$-stage
FAPM algorithm in terms of time performance. We also underline that  $|t^{(3)}_{f}-t^{(4)}_{f}|\leq\frac{2\pi}{\omega_0}$ holds for any pair of initial and
target states.

3.  From \textbf{Remarks} $1$-$4$ in Sect. 3, we conclude that transition time can be improved on the whole if frequencies of dynamical
radio-frequency(RF) field are permitted to vary continuously within some finite interval $[\omega_0-\omega^{-}_b,\omega_0+\omega^{+}_b]$ with
$\min\{\omega^{+}_{b},\omega^{-}_{b}\}\geq\omega^{max}_1=50kHz$ and $\omega_0=500MHz$. In fact, the transition time scale with frequency modulation is
about $\frac{\pi}{\omega^{max}_1}$ whereas the transition time scale without frequency modulation is about $\frac{4\pi}{\omega^{max}_1}$ if
$\min\{\omega^{+}_{b},\omega^{-}_{b}\}\geq\omega^{max}_1$ and $\omega_0\gg{\omega^{max}_1}$.  This implies that the transition time scale with frequency
modulation is about $1/4$ of that without frequency modulation for the worst case.

4. To explore  whether or not $1$-stage FAPM is  always better than $1$-stage APM in terms of time performance, we will carry on the following
calculations:

(i)  Let $\theta_0=\frac{3\pi}{4}$, $\phi_0=\frac{5\pi}{4}$, $\theta_f=\frac{\pi}{4}$ and $\phi_f=\frac{\pi}{4}$.  Since $\theta_f-\theta_0<0$, we have
\begin{equation}
\label{f-1}
k_{2}=\min\{k\in{Z^{+}}|k\geq\frac{(4\pi+\theta_f-\theta_0)\omega_0}{2\pi{\omega^{max}_1}}+1+\frac{\phi_{f}-\phi_0}{{2\pi}}\}=17501
\end{equation}
and
\begin{equation}
\label{f-2}
k_{4}=\min\{k\in{Z^{+}}|k\geq{R_{\omega}}+1+R_{B}\}=5001.
\end{equation}
Thus  $t^{(2)}_f-t_0=\frac{35000.5\pi}{\omega_{0}}=2.198\times10^{-4}s$ whereas $t^{(4)}_f-t_0=\frac{10001\pi}{\omega_{0}}=6.28\times10^{-5}s$.

(ii) Let $\theta_0=\frac{\pi}{4}$, $\phi_0=\frac{\pi}{4}$, $\theta_f=\frac{3\pi}{4}$ and  $\phi_f=\frac{5\pi}{4}$.  From $1$-stage FAPM algorithm, we have
$k_{4}=5001$ and $t^{(4)}_f-t_0=\frac{10001\pi}{\omega_{0}}\simeq6.28\times10^{-5}s$. This is in contrast with the observation that
\begin{equation}
\label{f-3}
k_{2}=\min\{k\in{Z^{+}}|k\geq\frac{(\theta_f-\theta_0)\omega_0}{2\pi{\omega^{max}_1}}+\frac{1}{4}+\frac{\phi_{f}-\phi_0}{{2\pi}}\}=2501
\end{equation}
 and $t^{2}_f-t_0=\frac{5001\cdot\pi}{\omega_{0}}=3.14\times10^{-5}s$. This implies that $1$-stage APM is better than $1$-stage FAPM for some pair of
 initial and target states in terms of time performance.

Therefore, neither algorithm is always better than the other in time performance considering different pairs of initial and target states. Although APM methods are studied under the assumption that $\omega_{rf}=\omega_0$ is a constant, by setting $\omega_{rf}=\omega_0$, a APM algorithm can also be used for the control of a parameters adjustable NMR system. As a result, in such a system the combination of 1-stage APM and 1-stage FAPM is feasible, and this hybrid modulation algorithm can be better than all 4 kinds of algorithms discussed in terms of time performance as the aforementioned calculation suggested. The hybrid modulation algorithm based on 1-stage FAPM and 1-stage APM algorithms is proposed in Table \ref{tab1}.
\begin{table}
\centering
\caption{Hybrid Modulation Algorithm} \label{tab1}
\begin{tabular}{ll}
\hline
\hline
C1: &Initial and target conditions $\theta_{0}$, $\phi_{0}$, $\theta_{f}$, $\phi_{f}$\\
\hline
C2:& Physical conditions $\omega_0$, $\omega^{max}_1$, $\omega^{-}_b$, $\omega^{+}_b$\\
\hline
C3:& Further Assumption $\min\{\omega^{max}_1,\omega^{+}_{b},\omega^{-}_{b}\}=\omega^{max}_1$\\
\hline
1 & Apply $1$-stage FAMP algorithm in Sect. 3.4 to calculate $\phi^{4}_{k}$.\\
2 &  Apply $1$-stage AMP algorithm in Sect. 3.2 to calculate $\phi^{2}_{k}$.\\
3  & if $\phi^{(2)}_{k}>\phi^{(4)}_{k}$, go to next step;\\
  & else, go to step 5.\\
4 & Apply $1$-stage FAMP algorithm in Sect. 3.4 to obtain \\
   &  the designed parameters $\omega^{(4)}_1$,  $\omega^{(4)}_{rf}$,  $\varphi^{(4)}_1$ and $t^{(4)}_f$,  go to  6.\\
5 &  Apply $1$-stage AMP algorithm in  to obtain \\
   & the designed parameters $\omega^{(2)}_1$,  $\varphi^{(2)}_1$, and $t^{(2)}_f$.\\
6 & End of the hybrid modulation algorithm.\\
\hline
\end{tabular}
\end{table}

5.  Some  may argue that the hybrid algorithm given in Table \ref{tab1} is quite complicated. We would like to further investigate the time scale of
transition time for FAPM and APM algorithms by  approximation.  For  nuclear spin systems,  we have $\omega_{0}/\omega_{1}^{max}\gg1$ and
$\omega_{0}/\min(\omega^{+}_{b},\omega^{-}_{b})\gg1$.  Therefore, $k_{2}$ and $k_{4}$ can be approximately  estimated by
\begin{equation}\label{simplifiedkap}
k_{2}\approx{k^{'}_{2}}=\left\{\begin{array}{ll}
\min\{k\in Z^{+}{\mid}k\geq\frac{(4\pi+\theta_{f}-\theta_{0})\omega_{0}}{2\pi\omega_{1}^{max}}\} &\  if \  \theta_f-\theta_0<0\\
\min\{k\in Z^{+}{\mid}k\geq\frac{(\theta_{f}-\theta_{0})\omega_{0}}{2\pi\omega_{1}^{max}}\} &\  if \  \theta_f-\theta_0\geq0
\end{array}\right.
\end{equation}
and
\begin{equation}\label{simplifiedkfap}
  k_{4}\approx{k^{'}_{4}}=\min\{k\in Z^{+}\mid k\geq
  \max(\frac{\omega_{0}|cos\frac{\theta_{0}+\theta_{f}}{2}|}{2\min(\omega^{+}_{b},\omega^{-}_{b})},\frac{\omega_{0}sin\frac{\theta_{0}+\theta_{f}}{2}}{2\omega^{max}_{1}})\}.
\end{equation}
This implies
\begin{equation}\label{tap}
  t_{f}^{(2)}-t_0=\frac{\phi_{k}^{(2)}}{\omega_{0}}\approx\frac{2 k_{2}\pi}{\omega_{0}}{\approx}\frac{2 k^{'}_{2}\pi}{\omega_{0}}=\tilde{t}_{f}^{(2)}-t_0
\end{equation}
and
\begin{equation}\label{tfap}
  t_{f}^{(4)}-t_0=\frac{\phi_{k}^{(4)}}{\omega_{0}}\approx\frac{2 k_{4}\pi}{\omega_{0}}\approx\frac{2 k^{'}_{4}\pi}{\omega_{0}}=\tilde{t}_{f}^{(4)}-t_0.
\end{equation}

Further calculations imply that
\begin{equation}\label{delta}
\Delta_{2}=|t_{f}^{(2)}-\tilde{t}_{f}^{(2)}|\leq\frac{4\pi}{\omega_{0}};\Delta_{4}=|t_{f}^{(4)}-\tilde{t}_{f}^{(4)}|\leq\frac{6\pi}{\omega_{0}}.
\end{equation}
Since $\omega_{0}\geq500MHz$, we have $\Delta_{2}\leq10^{-7}s$  and $\Delta_{4}\leq10^{-7}s$.

\begin{figure}
  \centering
  \includegraphics[width=\textwidth]{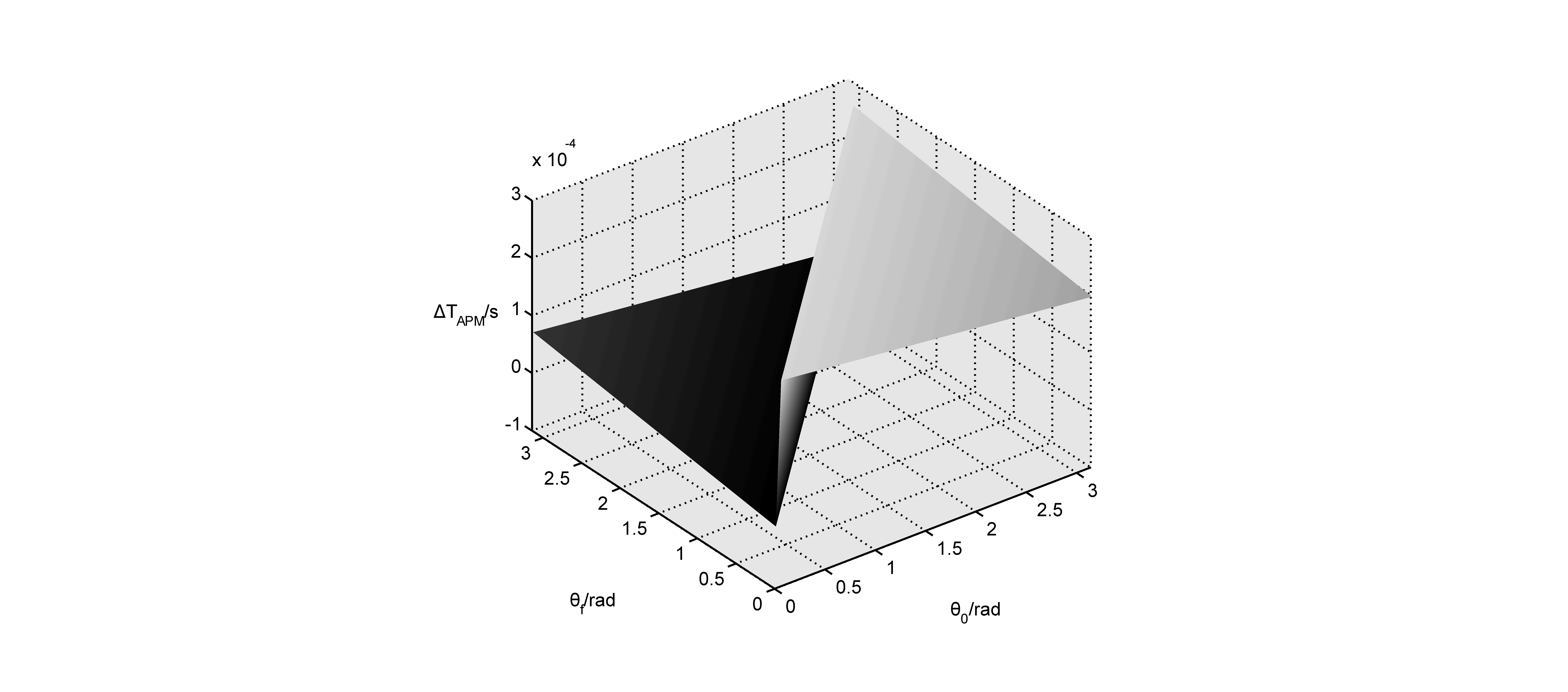}\\
  \caption{${\tilde{t}}^{(2)}_{f}-t_0$ when $\omega_{0}=500MHz$ and $\min\{\omega^{max}_1,\omega^{+}_{b},\omega^{-}_{b}\}=50kHz$}\label{ap}
\end{figure}

\begin{figure}
  \centering
  \includegraphics[width=\textwidth]{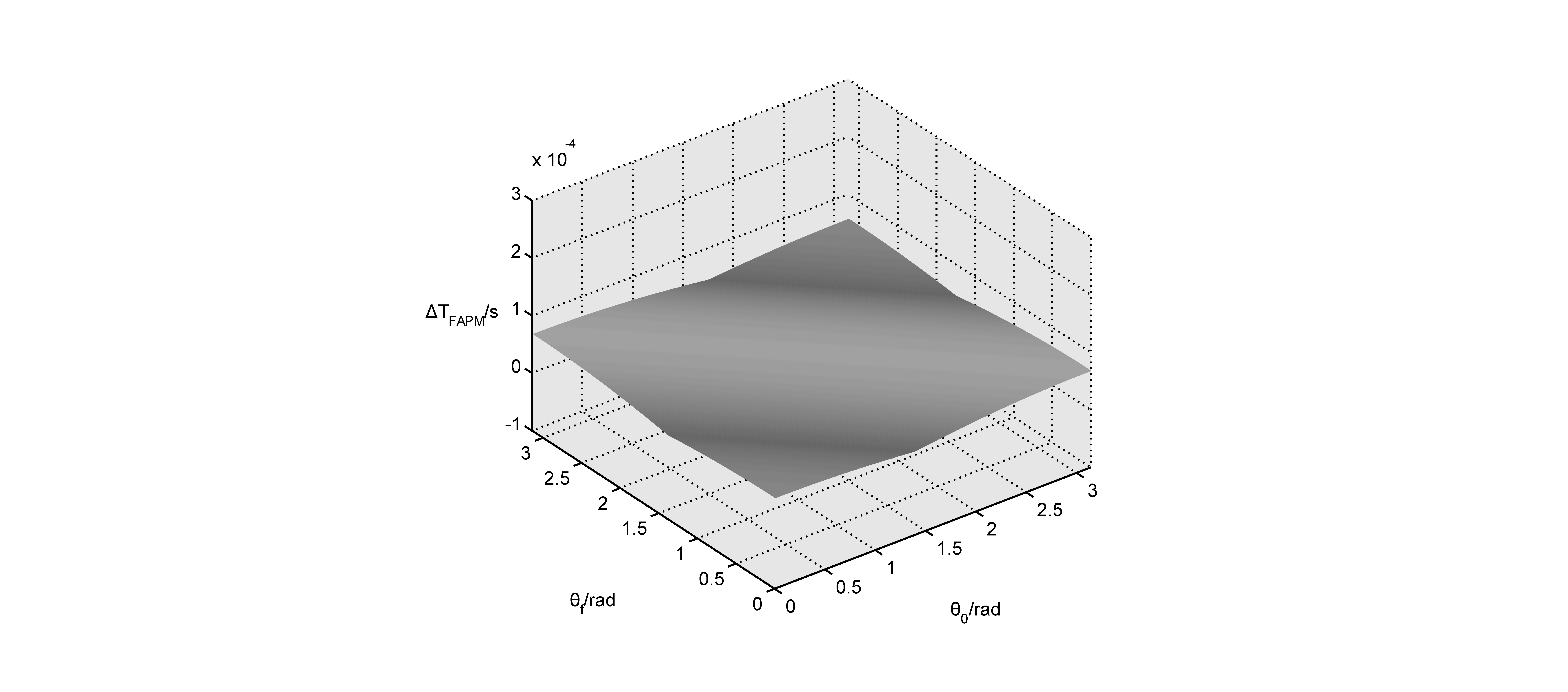}\\
  \caption{${\tilde{t}}^{(4)}_{f}-t_0$ when $\omega_{0}=500MHz$ and $\min\{\omega^{max}_1,\omega^{+}_{b},\omega^{-}_{b}\}=50kHz$}\label{fap}
\end{figure}

\begin{figure}
  \centering
  \includegraphics[width=\textwidth]{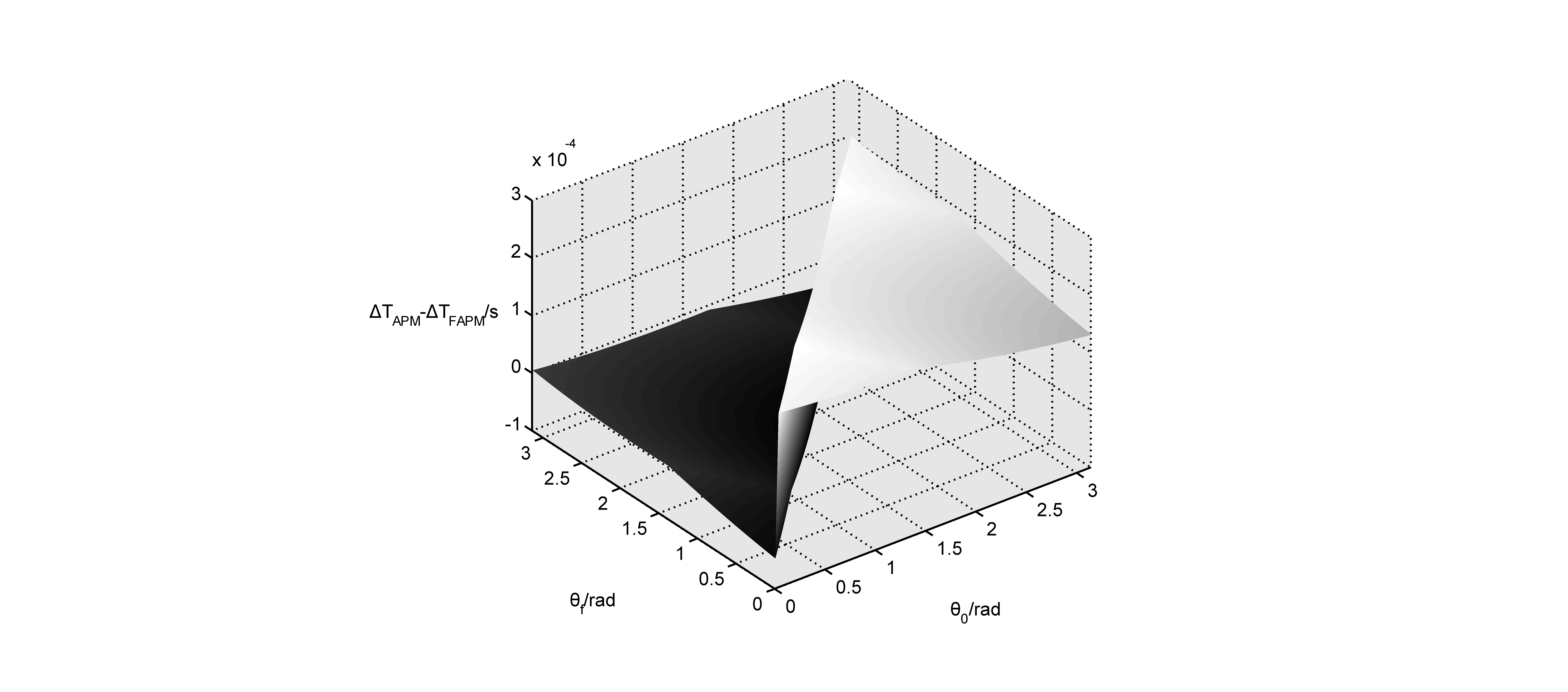}\\
  \caption{$\tilde{t}^{(4)}_{f}-\tilde{t}^{(2)}_{f}$ when $\omega_{0}=500MHz$ and
  $\min\{\omega^{max}_1,\omega^{+}_{b},\omega^{-}_{b}\}=50kHz$}\label{cmpapfap}
\end{figure}

Thus we can rather accurately estimate time performance $t_f^{(2)}-t_0$ and $t_f^{(4)}-t_0$ by calculating $k_2^{'}$ and $k_4^{'}$ respectively. It should be underlined that $k_2^{'}$ and $k_4^{'}$ are only functions of $\theta_0$ and $\theta_f$, and so are $t _f^{(2)}-t_0$ and $t _f^{(4)}-t_0$. To obtain an intuitive idea about the transition time scale for 1-stage APM and FAPM algorithms, we plot $\Delta T_{APM}=\tilde{t}_f^{(2)}-t_0$ and $\Delta T_{FAPM}=\tilde{t} _f^{(4)}-t_0$ in Fig. \ref{ap} and Fig. \ref{fap}, respectively. Furthermore, in order to obtain a picture about the difference between 1-stage APM algorithm and 1-stage FAPM algorithm, we also plot $\Delta T_{APM}-\Delta T_{FAPM}$ in Fig. \ref{cmpapfap}.

The calculation result in Fig. 3 corresponds to the simplified hybrid algorithm well. To plot these 3 figures, parameters of initial and target states, $\phi _0$ and $\phi _f$, have been given different values in attempts to discover how their values will influence the transition time. Nevertheless, the differences in transition time caused by the variance in $\phi _0$ and $\phi _f$ are less than $10^{-7}$ s with $\omega _0=500MHz$, and can hardly be distinguished in the figures. Such results implies us that a simplified hybrid modulation is feasible. Here in the figures given, we have set $\phi _0=\phi _f=0$.

6.  Let $\min\{\omega^{+}_{b},\omega^{-}_{b}\}\geq\omega^{max}_1$,  we have the following observations:
$k^{'}_{2}\leq k^{'}_{4}$ when $\theta_{f}-\theta_{0}\geq 0$; and $k^{'}_{2}>k^{'}_{4}$ when $\theta_{f}-\theta_{0}<0$.
Based on the aforementioned discussion, it is revealed  that one can approximately compare $t_{f}^{(2)}-t_0$  with $t_{f}^{(4)}-t_0$ just by calculating
$\theta_{f}-\theta_{0}$. Therefore a simplified hybrid algorithm is proposed in Table \ref{tab2} without comparing $t_{f}^{(2)}-t_0$  with
$t_{f}^{(4)}-t_0$.
\begin{table}
\centering
\caption{Simplified Hybrid Modulation Algorithm} \label{tab2}
\begin{tabular}{ll}
\hline
\hline
C1: &Initial and target conditions $\theta_{0}$, $\phi_{0}$, $\theta_{f}$, $\phi_{f}$\\
\hline
C2:& Physical conditions $\omega_0$, $\omega^{max}_1$, $\omega^{-}_b$, $\omega^{+}_b$\\
\hline
C3:& Further Assumption $\min\{\omega^{max}_1,\omega^{+}_{b},\omega^{-}_{b}\}=\omega^{max}_1$\\
\hline
1  & if $\theta_{0}>\theta_{f}$, go to next step;\\
  & else, go to step 3\\
2 &  Apply $1$-stage FAMP algorithm in Sect. 3.4 to obtain \\
   &  the designed parameters $\omega^{(4)}_1$,  $\omega^{(4)}_{rf}$,  $\varphi^{(4)}_1$ and $t^{(4)}_f$, \\
   &  and  go to  4.\\
3 &  Apply $1$-stage AMP algorithm in Sect. 3.2 to obtain\\
   &  the designed parameters $\omega^{(2)}_1$,  $\varphi^{(2)}_1$, and $t^{(2)}_f$,\\
4 & End of simplified hybrid algorithm.\\
\hline
\end{tabular}
\end{table}

The error between the hybrid modulation algorithm and its simplified version is nonzero only when (i)  $t^{(2)}_f-t^{(4)}_f>0$ but
$\tilde{k}_2<\tilde{k}_4$; or (ii) $t^{(2)}_f-t^{(4)}_f<0$ but $\tilde{k}_2>\tilde{k}_4$.
Notice also that  $t^{(2)}_f-t^{(4)}_{f}=\frac{2(k_2-k_4)\pi-\pi\cos\frac{\theta_0+\theta_f}{2}}{\omega_0}$
and $k_2-k_4=(k_2-\tilde{k}_2)+(\tilde{k}_2-\tilde{k}_4)-(k_4-\tilde{k}_4)$. Therefore, we conclude from Eq. (\ref{delta}) that the error between two
hybrid modulation algorithms can be estimated by
\begin{equation}
\label{d}
|t^{(2)}_f-t^{(4)}_{f}|\leq\Delta_2+\Delta_4+\frac{|\pi\cos\frac{\theta_0+\theta_f}{2}|}{\omega_0}\leq\frac{11\pi}{\omega_0}
\end{equation}

\begin{figure}
  \centering
  \includegraphics[width=\textwidth]{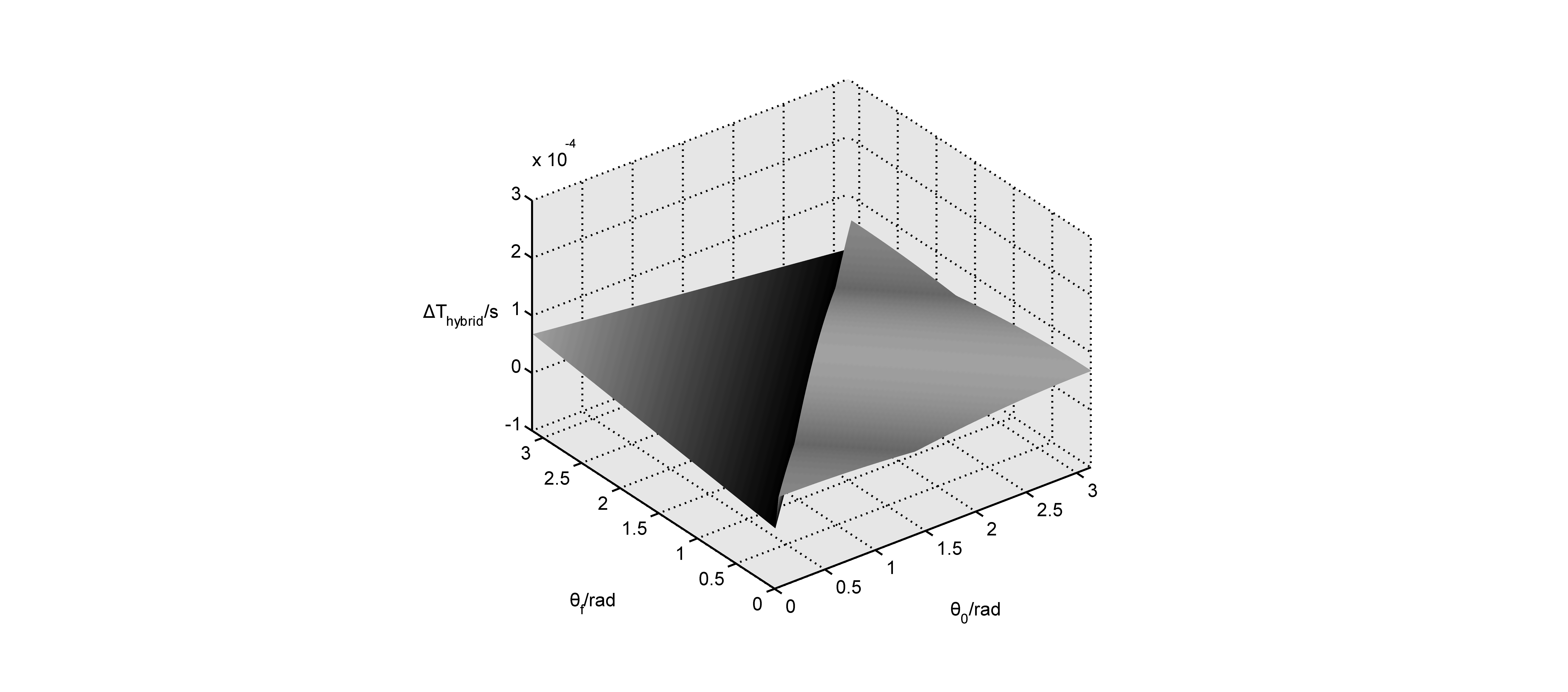}\\
  \caption{$\min\{{\tilde{t}}^{(2)}_{f},{\tilde{t}}^{(4)}_{f}\}-t_0$ when $\omega_{0}=500MHz$ and
  $\min\{\omega^{max}_1,\omega^{+}_{b},\omega^{-}_{b}\}=50kHz$}\label{hyp}
\end{figure}

In order to further obtain the intuitive ideas about transition time scale for the simplified hybrid modulation algorithm, we  plot
$\min\{{\tilde{t}}^{(2)}_{f},{\tilde{t}}^{(4)}_{f}\}-t_0$  in Fig. \ref{hyp}. If $\omega_1^{max}\leq min\{\omega_b^{+},\omega_b^{-}\}$, then the transition time scale for the simplified hybrid algorithm is estimated by the following equation:
\begin{equation}
\label{deltat}
t_f^{sh}-t_0\simeq \left\{\begin{array}{lc}
                            \frac{\pi}{\omega_1^{max}} & if \theta_f - \theta_0 \\
                            \frac{\theta_f-\theta_0}{\omega_1^{max}} & if \theta_f-\theta_0
                          \end{array}
\right..
\end{equation}

7. With the algorithms given in Table \ref{tab1} and \ref{tab2}, we are able to manage the state transition in a relative short time period. But one should still be aware of that these aforementioned algorithms are not optimal methods for any pair of initial and target states in terms of time performance. In a general amplitude-phase-frequency adjustable NMR control system, to get such an optimal algorithm, one should consider both the time spend and the consumptions in other aspects, such as energy, transition error and amplitude of magnetic field, and try to get the result by solving some cost functions. Nevertheless, the analytical solutions to this kind of optimal problems are either in a complicated expression or too difficult to get in general.

In this paper, we try to avoid dealing with such problems by adding some more constraints. Here we require the state transition should be exact, i.e. $|\psi(t_f)\rangle=|\psi_f\rangle$, and that the deployment of $\omega_1$, $\omega_{rf}$ and $\phi_1$ should obey aforementioned APM or FAPM algorithms. Under such constraints, the aforementioned time optimization should be regard as a special case of the general optimal control problem. In other words, it becomes an optimal control problem with constrains on exact realization in state transition and finite choices in parameters determination.

As a result, the hybrid algorithm and its simplified version are different from the general case of optimal algo-rithms in terms of time performance. The time consumption of hybrid algorithms is the shortest among the APM and FAPM algorithms discussed. From this point of view, hybrid algorithms are indeed optimal algorithms. On the other hand, extra constraints of exact state transition and limits to transition methods is added. These constraints make the algorithms no longer an overall optimal process. According to such reasons, we emphasize that the hybrid algorithm and its simplified version are 2 instances of sub-optimal algorithms.

\section{Conclusion}

In summary, it is revealed in this research that transition time of steering nuclear spin states can be improved on the whole if exact frequency modulation is available. It is also exemplified that the transition time with frequency modulation is about 1/4 of that without frequency modulation. For any pair of initial and target states, 1-stage APM algorithm is better than 3-stage APM algorithm and 1-stage FAPM algorithm is better than 2-stage FAPM algorithm in terms of transition time performance. However, 1-stage FAPM algorithm is not always better than 1-stage APM algorithm in time performance for any pair of initial and target states. Based on a careful analysis, the hybrid scheme of 1-stage FAPM and 1-stage APM algorithms is proposed in Table 1, and it is better than 4 kinds of modulation algorithms in terms of time performance. The simplified hybrid scheme is further presented in Table 2 to reduce half of the computational burden, whereas the error between the hybrid scheme and simplified hybrid scheme is at most $11\pi$/$\omega_0$. Neither the hybrid modulation scheme nor its simplified version is optimal modulation method in terms of the Bloch parameters of initial and target states, but the hybrid modulation algorithm and its simplified version are constructive sub-optimal algorithms in terms of time performance with the constrain of transferring the state exactly and choosing control method between APM and FAPM algorithm. It is recently reported by Li et al.\cite{Li} that sinusoidal modulation can be used for manipulating a superconducting qubit, and this enhances our believes that frequency modulation should be a promising technique for manipulating qubits.

\section{ACKNOWLEDGMENTS}
M. Zhang is particularly grateful to Dr. H. D. Yuan, Dr. R. Wu and Dr. B. Qi for their constructive discussions.  This work was supported by the Program
for National Natural Science Foundation of P. R. China (Grant Nos.  61273202, 61134008 and 11074307).

\section{Appendix: Lemmas}
To study modulation algorithms, we introduce  the following lemmas:

\textbf{Lemma 1}: For any real function $f(t)$ of time $t$, we have
\begin{equation}
\label{Lemma1}
exp\{-if(t)S_z\}(\cos{f(t)}S_x-\sin{f(t)}S_y)exp\{if(t)S_z\}=S_x.
\end{equation}

\textbf{Proof:} This lemma is proved by showing that
\begin{equation}
\label{Lemma1-1}
\left(\begin{array}{cc}
e^{\frac{-if(t)}{2}} &0 \\
0 & e^{\frac{if(t)}{2}}
\end{array}\right)\cdot
\left(\begin{array}{cc}
0 &\frac{1}{2}e^{{if(t)}} \\
\frac{1}{2}e^{{-if(t)}} & 0
\end{array}\right)\cdot
\left(\begin{array}{cc}
e^{\frac{if(t)}{2}} &0 \\
0 & e^{-\frac{if(t)}{2}}
\end{array}\right)=\left(\begin{array}{cc}
0 &\frac{1}{2} \\
\frac{1}{2} & 0
\end{array}\right).
\end{equation}

\textbf{Lemma 2}: The solution of
\begin{equation}
\label{Lemma2-1}
\frac{d}{dt}|\psi(t)\rangle=
i\{\omega_0S_z+\omega_1[S_x\cos[\omega_0(t-\tau)+\varphi_1]-S_y\sin[\omega_0(t-\tau)+\varphi_1]]\}|\psi(t)\rangle
\end{equation}
with the initial state $|\psi(\tau)\rangle$ at time $\tau$ is given by
\begin{equation}
\label{Lemma2-2}|\psi(t)\rangle=
exp\{i[\omega_0(t-\tau)+\varphi_1]S_z\}exp\{i\omega_1(t-\tau)S_x\}exp\{-i\varphi_1{S_z}\}|\psi(\tau)\rangle.
\end{equation}

\textbf{Proof:}  Denote $|\phi(t)\rangle=exp\{-i[\omega_0(t-\tau)+\varphi_1]S_z\}|\psi(t)\rangle$.

From Lemma 1 with $f(t)=\omega_0(t-\tau)+\varphi_1$, we have
\begin{equation}
\label{Lemma2-3}
\frac{d}{dt}|\phi(t)\rangle=i\omega_1S_x|\phi(t)\rangle.
\end{equation}
After some calculations, we obtain Eq. (\ref{Lemma2-2}) from the observations that
$|\psi(t)\rangle=exp\{i[\omega_0(t-\tau)+\varphi_1]S_z\}|\phi(t)\rangle$ and $|\phi(\tau)\rangle=exp\{-i\varphi_1{S_z}\}|\psi(\tau)\rangle$.

\textbf{Lemma 3}: The solution of
\begin{equation}
\label{Lemma3-1}
\frac{d}{dt}|\psi(t)\rangle=
i\{\omega_0S_z+\omega_1[S_x\cos[\omega_{rf}(t-\tau)+\varphi_1]-S_y\sin[\omega_{rf}(t-\tau)+\varphi_1]]\}|\psi(t)\rangle
\end{equation}
with the initial state $|\psi(\tau)\rangle$  at time $\tau$ is given by
\begin{equation}
\label{Lemma3-2}
|\psi(t)\rangle=
e^{i[\omega_{rf}(t-\tau)+\varphi_1]S_z}exp\{iR_u(t-\tau)S_{\theta_u}\}e^{-i\varphi_1{S_z}}|\psi(\tau)\rangle
\end{equation}
where  $R_u=\sqrt{(\omega_0-\omega_{rf})^{2}+\omega^{2}_1}$,  $\cos\theta_u=\frac{\omega_0-\omega_{rf}}{R_u}$ and $\sin\theta_u=\frac{\omega_1}{R_u}$ and
$S_{\theta_u}=\cos\theta_uS_z+\sin\theta_uS_x$.

\textbf{Proof:} Denote $|\phi(t)\rangle=exp\{-i[\omega_{rf}(t-\tau)+\varphi_1]S_z\}|\psi(t)\rangle$.

From Lemma 1 with $f(t)=\omega_{rf}(t-\tau)+\varphi_1$, we have
\begin{equation}
\label{Lemma3-3} \frac{d}{dt}|\phi(t)\rangle=i[(\omega_0-\omega_{rf})S_z+\omega_1S_x]|\phi(t)\rangle.
\end{equation}
 After some calculations, we obtain Eq. (\ref{Lemma3-2}) from  the  following observations that $(\omega_0-\omega_{rf})S_z+\omega_1S_x={R_u}S_{\theta_u}$,
 $|\psi(t)\rangle=e^{i[\omega_{rf}(t-\tau)+\varphi_1]S_z}|\phi(t)\rangle$ and $|\phi(\tau)\rangle=e^{-i\varphi_1{S_z}}|\psi(\tau)\rangle$.

\textbf{Lemma 4}: For $\forall{g}\in{R}$, we have
\begin{equation}
\label{Lemma4-1}
exp\{igS_x\}(\cos\frac{\theta_0}{2}|\uparrow\rangle+i\sin\frac{\theta_0}{2}|\downarrow\rangle)=\cos\frac{g+\theta_0}{2}|\uparrow\rangle+i\sin\frac{g+\theta_0}{2}|\downarrow\rangle.
\end{equation}

\textbf{Proof:}  Note that
\begin{equation}
\label{Lemma4-3}
\left(\begin{array}{cc}
\frac{1}{\sqrt{2}} &\frac{1}{\sqrt{2}} \\
\frac{1}{\sqrt{2}} & -\frac{1}{\sqrt{2}}
\end{array}\right)\cdot{S_z}\cdot
\left(\begin{array}{cc}
\frac{1}{\sqrt{2}} &\frac{1}{\sqrt{2}} \\
\frac{1}{\sqrt{2}} & -\frac{1}{\sqrt{2}}
\end{array}\right)=S_x.
\end{equation}
Thus this lemma is proved from the following observation
\begin{equation}
\label{Lemma4-2}
\left(\begin{array}{cc}
\frac{1}{\sqrt{2}} &\frac{1}{\sqrt{2}} \\
\frac{1}{\sqrt{2}} & -\frac{1}{\sqrt{2}}
\end{array}\right)\cdot\left(\begin{array}{cc}
e^{i\frac{g}{2}} &0 \\
0 & e^{-i\frac{g}{2}}
\end{array}\right)\cdot
\left(\begin{array}{cc}
\frac{1}{\sqrt{2}} &\frac{1}{\sqrt{2}} \\
\frac{1}{\sqrt{2}} & -\frac{1}{\sqrt{2}}
\end{array}\right)\cdot
\left(\begin{array}{c}
\cos\frac{\theta_0}{2} \\
i\sin\frac{\theta_0}{2}
\end{array}\right)=\left(\begin{array}{c}
\cos\frac{\theta_0+g}{2} \\
i\sin\frac{\theta_0+g}{2}
\end{array}\right)
\end{equation}

\textbf{Lemma 5}:
\begin{equation}
\label{Lemma5-1}
exp\{i\pi{S_{\theta_u}}\}(\cos\frac{\theta_0}{2}|\uparrow\rangle+\sin\frac{\theta_0}{2}|\downarrow\rangle)=\cos(\theta_u-\frac{\theta_0}{2})|\uparrow\rangle+\sin(\theta_u-\frac{\theta_0}{2})|\downarrow\rangle
\end{equation}
\textbf{Proof:}  This lemma is proved from the following observations that
\begin{equation}
\label{Lemma5-2}
U_{\theta_u}\cdot{S_{z}}\cdot
U_{\theta_u}=\cos\theta_uS_z+\sin\theta_uS_x=S_{\theta_u}
\end{equation}
and
\begin{equation}
\label{Lemma5-3}
U_{\theta_u}\cdot\left(\begin{array}{cc}
e^{i\frac{\pi}{2}} &0 \\
0 & e^{-i\frac{\pi}{2}}
\end{array}\right)\cdot
U_{\theta_u}\cdot
\left(\begin{array}{c}
\cos\frac{\theta_0}{2} \\
\sin\frac{\theta_0}{2}
\end{array}\right)=i\left(\begin{array}{c}
\cos(\theta_u-\frac{\theta_0}{2}) \\
\sin(\theta_u-\frac{\theta_0}{2})
\end{array}\right)
\end{equation}
with
\begin{equation}
\label{Lemma5-4}
U_{\theta_u}=\left(\begin{array}{cc}
\cos\frac{\theta_u}{2} &\sin\frac{\theta_u}{{2}} \\
\sin\frac{\theta_u}{{2}} & -\cos\frac{\theta_u}{{2}}
\end{array}\right).
\end{equation}

%% References


%\bibliographystyle{model3a-num-names}

%% not want to use model3a-num-names.bst.


\end{document}